\documentclass[12pt]{article}
\usepackage{epsfig}
\usepackage{amsfonts}
\usepackage{amssymb}

\def\half{\frac{1}{2}}

\newcommand{\qhat}{\hat{q}_0}

\newcommand{\eqn}[1]{(\ref{#1})}
\newcommand{\be}{\begin{equation}}
\newcommand{\ee}{\end{equation}}
\newcommand{\ben}{\begin{displaymath}}
\newcommand{\een}{\end{displaymath}}
\newcommand{\bea}{\begin{eqnarray}}
\newcommand{\eea}{\end{eqnarray}}
\newcommand{\bean}{\begin{eqnarray*}}
\newcommand{\eean}{\end{eqnarray*}}
\newcommand{\nn}{\nonumber \\}
\newcommand{\ba}{\begin{array}}
\newcommand{\ea}{\end{array}}
\newcommand{\bi}{\begin{itemize}}
\newcommand{\ei}{\end{itemize}}

\newcommand{\reef}[1]{(\ref{#1})}

\newcommand{\mt}[1]{\textrm{\tiny #1}}

\def\G{\Gamma}

\def\e{\epsilon}

\def\otaula{\begin{tabular}}
\def\ctaula{\end{tabular}}

\def\bnum{\begin{enumerate}}
\def\enum{\end{enumerate}}

\def\CR{\mathbb{R}}
\def\CM{\mathcal{M}}

\def\8M{$\CM_8$}

\def\be{\begin{equation}}
\def\ee{\end{equation}}
\def\G{\Gamma}

\def\ei{e^{\underline{i}}}

\def\e1{e^{\underline{1}}}
\def\1u{\underline{1}}
\def\2u{\underline{2}}

\def\0u{\underline{0}}
\def\e{\epsilon}
\def\target{$\CR^{1,1}\times \mathcal{M}_8$ }
\def\target2{$\CR^{1,1}\times \mathcal{M}_8$,}
\def\9G{\G_{\underline{9}}}

\newcommand{\cale}{\mbox{${\cal E}$}}

\newcommand{\calk}{\mbox{${\cal K}$}}

\newcommand{\calm}{\mbox{${\cal M}$}}
\newcommand{\caln}{\mbox{${\cal N}$}}

\newcommand{\sac}{\, , \qquad}
\newcommand{\eg}{{\it e.g. }}
\newcommand{\ie}{{\it i.e. }}

\begin{document}

\begin{titlepage}
\begin{flushright}
hep-th/0506110
\end{flushright}
\vskip 1.5in
\begin{center}
{\bf\Large{Oscillator Level for Black Holes and Black Rings}}
 \vskip 0.5in {\bf Roberto Emparan$^{a,b}$ and David Mateos$^{c}$}
 \vskip 0.3in {\small
{\textit{$^{a}$ Instituci\'o Catalana de Recerca i Estudis
Avan\c cats (ICREA),}\\
{\it $^{b}$ Departament de F{\'\i}sica Fonamental, and}\\
{\it C.E.R. en Astrof\'{\i}sica, F\'{\i}sica de Part\'{\i}cules i Cosmologia,}\\
{\it Universitat de Barcelona, Diagonal 647, E-08028 Barcelona,
Spain} }}
\\
{\small{ \textit{ $^c$Perimeter Institute for Theoretical Physics,
Waterloo, Ontario N2L 2Y5, Canada}}}

\end{center}
\vskip 0.5in

\baselineskip 16pt
\date{}

\begin{abstract}

Microscopic calculations of the Bekenstein-Hawking entropy of
supersymmetric black holes in string theory are typically based on the
application to a dual 2D CFT of Cardy's formula,
$S=2\pi \sqrt{c \, \hat{q}_0/6}$, where $c$ is the central
charge and $\hat{q}_0$ is the oscillator level. In the CFT,
$\qhat$ is non-trivially related to the total momentum.
We identify a Komar integral that equals
$\hat{q}_0$ when evaluated at the horizon, and the total momentum
when evaluated at asymptotic infinity, thus providing a
gravitational dual of the CFT result.

\end{abstract}
\end{titlepage}
\vfill\eject

\section{Introduction}

Our current understanding of the Bekenstein-Hawking entropy
formula within string theory takes its most precise form for the
supersymmetric black holes that are described in terms of a dual
two-dimensional (2D) conformal field theory (CFT). Typically, this
CFT preserves supersymmetry in, say, the right-moving sector, and
carries excitations in its left-moving sector up to oscillator
level
$L_0=\hat{q}_0$. A general argument then gives the entropy of the
system as
\begin{equation}
\label{cardy}
  S=2\pi\sqrt{\frac{c\, \hat{q}_0}{6}}\,,
\end{equation}
where $c$ is the central charge of the left-moving sector. An important
question is how $c$ and $\hat{q}_0$ are encoded in a given supergravity
black hole solution, so that \reef{cardy} reproduces precisely the
Bekenstein-Hawking entropy. A simple prescription for obtaining $c$ is
provided by the AdS/CFT correspondence, which relates it to the
effective cosmological constant of the near-horizon AdS$_3$ factor of
the solution. We shall show that $\qhat$ is directly
computed by a simple Komar integral at the black hole horizon. The same
integral, taken on a surface at infinity, gives the total momentum of
the CFT.

The CFT calculation of the oscillator level may be rather
involved, so it might be surprising that a simple general formula
computes the relation between the total momentum and
$\hat{q}_0$ on the gravity side. This result is also peculiar in
that it identifies a microscopic parameter with a quantity
measured at the horizon, instead of a conserved charge at
infinity. In fact, the earliest microscopic calculations
identified
$c$ and $\hat{q}_0$ using exclusively the asymptotic conserved
charges of the supergravity solution. In the original example of
\cite{SV} the presence of D1- and D5-brane charges
$Q_1$ and $Q_5$ leads to consideration of a D1-D5 bound state that
at low energies is described by an
$\caln=(4,4)$ CFT with central charge $c= 6 Q_1 Q_5$.
If the ADM momentum $P$ is identified with the oscillator number
$\hat{q}_0$, then the Bekenstein-Hawking entropy is exactly
reproduced.

In general, however, the oscillator number $\hat{q}_0$ is not
equal to the total momentum $P$ measured at infinity. For example,
some of the momentum may be carried by excitations that also carry
other conserved charges. The condition that the black hole carries
some specific values of these charges reduces the number of ways
in which the total momentum $P$ may be distributed amongst the
oscillators, so $\qhat < P$. A well-known example is the rotating
five-dimensional BMPV black hole \cite{BMPV}, for which some of
the left-moving fermionic momentum modes carry R-charge
corresponding to spacetime angular momentum.

A more subtle and intriguing example is provided by the recently
discovered supersymmetric black ring \cite{EEMR1} (see also
\cite{GG1,EEMR2,BW,GG2} for extensions), a five-dimensional black hole
with horizon topology $S^1 \times S^2$. A microscopic account of
its entropy, using the 2D CFT living on the intersection of three
M5-branes \cite{MSW}, has been given in
\cite{CGMS,BK}. Since the M5-branes intersect in
${\mathbb R}^4$ over the ring circle, which is contractible,
they carry no conserved charges. Instead, they give rise to three
dipoles, $q_i$, measured by flux integrals through spheres that
link the ring circle once. The configuration also includes three
stacks of orthogonal M2-branes that, from the microscopic
viewpoint, are described by fluxes on the worldvolume of the
M5-branes. These M2-branes carry three conserved charges, $Q_i$,
that can be measured at infinity.

One key difference between the microscopic description of the
black ring in \cite{CGMS,BK} and that of all other supersymmetric
black holes is that, in the black ring case, $c$ is not determined
by the conserved charges, $Q_i$, but by the non-conserved dipoles,
$q_i$.\footnote{A description within the CFT determined by the
conserved charges has been investigated in \cite{BK}.} The
momentum along the ring circle gives rise to an angular momentum,
$J$, at infinity. Again, this is not the same as $\hat{q}_0$,
which depends on $Q_i$ and $q_i$. In this case, the difference is
due to the fact that part of the momentum cannot be freely
distributed amongst the oscillators, since it arises from zero
modes associated to the M2-brane charges and from zero-point
contributions of the oscillators that shift the ground state
momentum to a non-zero value.

The black ring illustrates the fact that, in general, $c$ and
$\qhat$ cannot be determined exclusively in terms of the asymptotic
charges of the black hole solution. As mentioned above, however, a
universal prescription for the central charge is obtained by
considering the near-horizon geometry. Our proposal is a similar
prescription for the oscillator level. We argue that it can be
obtained as the value at the black hole horizon of an integral
that, at infinity, corresponds to an ADM conserved quantity.

The black hole solutions that admit a dual 2D CFT have a horizon
that extends along the spatial direction of the CFT (the
`effective string'). For a spherical black hole in $D$ dimensions,
this direction is only manifest after the black hole is uplifted
to a black string in $D+1$ dimensions; for black rings, it simply
corresponds to the direction along the $S^1$ of the ring. We
assume
that a Killing vector field $\ell$ generates translations along the
(compact) direction of this effective string, and is normalized so
that its orbits have period
$2\pi$. Let $S$ be a topological sphere that links this string
once, and denote also by $\ell$ the one-form dual to the Killing
vector. Then we consider the Komar integral
\begin{equation}
\label{komar}
\calk= \frac{1}{8G}\int_{S} *(\ell \wedge d\ell)\,,
\end{equation}
where $G$ is Newton's constant in the space where the black string
or ring lives. The reader might be more familiar with integrals of
the form $\int *d\ell$, suitable for black holes. In fact, if
$\ell$ is a Killing vector then one can easily show that
\be
\int_{S} *(\ell \wedge d\ell)=\frac{1}{2\pi}\int_{S\times S^1}
*d\ell\,,
\ee
where the $S^1$ is an orbit of $\ell$. The form \eqn{komar} is
more appropriate for objects which extend along the orbits of
$\ell$, such as black rings or strings \cite{TZ}.

The physical interpretation of the Komar integral is that it
measures the amount of momentum conjugate to $\ell$ in the volume
inside $S$ (see \eg \cite{Wald,TZ}). For example, if $S$ is a sphere
at infinity then $\calk$ equals the total (appropriately
quantized) momentum in the spacetime, namely the ADM linear
momentum for a black string, or the ADM angular momentum for a
black ring. In either case, $\calk_\infty$ corresponds to the
total momentum of the dual CFT.

We will show that, at the horizon, $\calk$ gives exactly the
oscillator level:
\begin{equation}
\label{main}
\calk_\mathrm{hor}=\hat{q}_0\,.
\end{equation}
In other words, the difference between the total momentum and the
oscillator level may be exactly identified with the amount of
momentum contained in the region outside the horizon. The same
integral \eqn{komar} was previously used in
\cite{RE} to compute the oscillator level of an extremal, but
non-supersymmetric, black ring.

The fact that the Komar integral at infinity measures the
corresponding ADM quantity is well-known. It may appear more
mysterious that the same integral at the horizon reproduces the
CFT oscillator level. However, we will show in section 2 that this
follows from simple, universal properties of the near-horizon
geometries of supersymmetric black holes that admit a dual 2D CFT
description. Instead, what is more surprising is that the same
Komar integral interpolates, on the gravity side, between two
quantities that in general differ from each other by a function of
the charges and dipoles whose direct calculation in the CFT may be
rather non-trivial.

There exist supersymmetric, string-like \cite{HM} or ring-like
\cite{BW,BWW} solutions with inhomogeneous,
singular horizons. These fail to be regular event horizons in the
sense that the metric cannot be analytically extended through
them, but a well-defined, finite area can still be ascribed to
them, hence also an entropy. The inhomogeneity arises from the
fact that these solutions are characterized by arbitrary functions
of the coordinate along the orbits of $\ell$, which implies that
$\ell$ is no longer an exact isometry. Asymptotically, $\ell$ is still a
Killing field, so the integral \eqn{komar} at infinity still
measures the total ADM charge. For any other surface of
integration, its interpretation as a measure of a conserved charge
inside a given volume is in principle lost. However, we will see
that its value at the inhomogeneous horizon still reproduces the
local entropy density.

In section \ref{examples} we explicitly verify the general result
for several relevant supersymmetric black hole solutions. In
particular, we consider the BMPV and the asymptotically flat black
ring examples in some detail because they nicely illustrate the
two main types of black hole horizons known to date. We also
briefly discuss the inhomogeneous solutions of \cite{BWW}. We
conclude in section \ref{discussion}.

\section{General argument}
\label{general}

Let us assume that the black hole admits a description in terms of
a dual theory which flows in the infrared to a 2D CFT. The AdS/CFT
correspondence then implies that the near-horizon geometry of the
black hole\footnote{See refs.~\cite{Reall, Gutowski} for a general
analysis of the near-horizon geometry of supersymmetric 5D black
holes.} is the direct product of the near-horizon geometry of the
extremal BTZ black hole and a compact space $\calm_{D-3}$, so the
metric takes the form
\be
\label{nearbtz}
ds^2 = \frac{4r_+}{l}r dt d\varphi +r_+^2d\varphi^2
+\frac{l^2}{4r^2}dr^2 +ds^2(\calm_{D-3})\,,
\ee
where $r_+$ and $l$ are the horizon radius and AdS$_3$ curvature
radius, respectively. Recall that the near-horizon geometry and
the decoupling limit geometry that is relevant in the AdS/CFT
context are in general different, as the former describes only the
infrared region of the latter. Since we wish to evaluate a Komar
integral at the horizon, it suffices to consider the near-horizon
geometry.

Observe that the first three-dimensional factor in
\reef{nearbtz} is {\em not} the same as the full geometry of the extremal BTZ
black hole,
\be\label{extbtz}
ds^2=-\frac{(\rho^2-r_+^2)^2}{l^2 \rho^2} \, d\tau^2+
\frac{l^2 \rho^2}{(\rho^2-r_+^2)^2} \, d\rho^2 +
\rho^2 \left(d\varphi +\frac{\rho^2-r_+^2}{l \rho^2} \, d\tau\right)^2
\ee
(written here in coordinates that corotate with the horizon) but only its
limit near the horizon at $\rho=r_+$, obtained by setting $\rho =r_++
\epsilon r$, $\tau=t/\epsilon$, and
taking $\epsilon\to 0$. The
distinction between the extremal BTZ geometry \reef{extbtz} and its
near-horizon limit \reef{nearbtz}
makes sense because, although the BTZ black hole is
locally equivalent to AdS$_3$, it is not a homogeneous space due
to global identifications. There is of course a
local coordinate change that takes one metric to the other, namely,
\be
r=\frac{\rho^2-r_+^2}{2r_+}\,,\qquad t=\tau +\frac{l}{2}
\, \varphi,
\ee
but making this map globally well defined requires a change in the
identifications of the coordinates --- in particular, it requires
periodically identifying $t\sim t+\pi l$, which introduces closed
timelike curves. Hence the metric \reef{nearbtz} is globally
inequivalent to the extremal BTZ metric \reef{extbtz}.

We therefore emphasize that we do not assume that the decoupling
limit contains a {\it full} extremal BTZ black hole \reef{extbtz}.
For example, the decoupling limit of the black ring of
\cite{EEMR1} does not contain a factor \reef{extbtz}, but it
does have
\reef{nearbtz} as its near-horizon geometry (see \cite{EEMR2} for a
detailed discussion).

The entropy associated to the horizon of \reef{nearbtz} is
\be
S= 2\pi r_+ \frac{V_{D-3}}{4 G} \,,
\label{BTZentropy}
\ee
where $V_{D-3}$ is the volume of $\calm_{D-3}$. The coordinates
$t$ and $\varphi$ parametrize  isometric directions that
correspond to the time and spatial directions of the dual infrared
CFT, respectively.
We are therefore interested in the Killing vector field
$\partial_\varphi$, whose dual one-form $\ell$ is
\be
\ell=r_+^2 d\varphi +\frac{2r_+}{l}r dt\,.
\ee
The associated Komar integral \reef{komar} gives
\be
\calk=\frac{V_{D-3}}{4 G}\frac{r_+^2}{l}\,.
\ee
If we now use the AdS/CFT relation $c=3l/2G_3$, together with
$G_3=G/V_{D-3}$, then we see that $\hat{q}_0=\calk$ is the
right oscillator number for the 2D CFT formula \reef{cardy} to
reproduce the Bekenstein-Hawking entropy \eqn{BTZentropy}.

This calculation concerns only the near-horizon geometry. In the
full asymptotically flat solution, $\calk_\infty$ will compute a
conserved ADM quantity associated to the total momentum that in
general will not be the same as the value $\hat{q}_0$ at the
horizon. For instance, in the cases of the BMPV black hole and the
black ring, the transverse three-sphere at infinity differs from
the three-sphere at the horizon by diffeomorphisms that involve
$t$ and $\varphi$ \cite{CL}. For BMPV, these are well-understood to be the
the gravitational duals of the spectral flow in the CFT
responsible for the difference between $P$ and $\hat{q}_0$. For
the black ring, however, the analogous gravitational
interpretation of the CFT calculation of $\hat{q}_0$ is less well
understood. From this viewpoint, the main observation in this
paper is that the integral \reef{komar} precisely computes this
quantity on the gravity side. In the next section we discuss
explicitly the most relevant examples.

\section{Examples}
\label{examples}

\subsubsection*{BMPV black hole}

This is a rotating black hole in five dimensions. As explained
in the introduction, we lift it to six dimensions in order to exhibit more
transparently its relation to the dual microscopic CFT. The metric
takes the form\footnote{The six-dimensional dilaton is
$e^\Phi=1$, so the string- and Einstein-frame metrics are the
same.}
\bea
\label{6dmet}
ds^2 &=& -H_\mt{P}^{-1} (H_1 H_5)^{-1/2} \, (dt + \omega)^2
+ H_\mt{P} (H_1 H_5)^{-1/2} \, [ dz + H_\mt{P}^{-1} (dt+ \omega) ]^2
+ (H_1 H_5)^{1/2} \, d{\bf x}_{\it 4}^2 \nonumber \\
&=& H_\mt{P} (H_1 H_5)^{-1/2} \, dz \, [ dz
+ 2H_\mt{P}^{-1} (dt+ \omega)]
+ (H_1 H_5)^{1/2} \, d{\bf x}_{\it 4}^2 \,,
\eea
where
\be
d{\bf x}_{\it 4}^2= d\rho^2 + \rho^2 (d\theta^2
+ \sin^2 \theta \, d\psi^2 + \cos^2 \theta \, d\phi^2) \,,
\ee
\be
\label{metfun}
H_\mt{P} = 1 + \frac{Q_\mt{P}}{\rho^2} \sac H_1= 1 +
\frac{Q_1}{\rho^2}
\sac H_5= 1 + \frac{Q_5}{\rho^2} \,,
\ee
and
\be
\omega =
\frac{2G}{\pi^2 R_z}\frac{J}{\rho^2} \left(\sin^2 \theta\, d\psi
+ \cos^2 \theta \, d\phi\right) \,.\ee
$R_z$ is the radius of the CFT direction, $z$, and $G$ is the
six-dimensional Newton's constant.
The charges $Q_1$ and $Q_5$ may be thought of as corresponding to
the numbers of constituent D1-branes and D5-branes in the system,
respectively, and satisfy
$Q_1Q_5=(2G/\pi^2)N_1 N_5$. The ADM momentum in the $z$-direction
is
\be
P=\frac{\pi^2 R_z}{2 G}Q_\mt{P}=\frac{N_\mt{P}}{R_z} \,,
\ee
where $N_\mt{P}$ is the integer number of momentum units. $J$ is
the integrally quantized angular momentum of the black hole. The
horizon lies at $\rho =0$. The entropy is given by equation
\eqn{cardy} with $c = 6 N_1 N_5$ and
\be
\hat{q}_0 = N_\mt{P} - \frac{J^2}{N_1 N_5} \,.
\ee

Thus we are interested in the Komar integral associated to the
Killing vector field $R_z\partial_z$, whose dual one-form is
\be
\ell = R_z H_\mt{P} (H_1 H_5)^{-1/2} \, [ dz + H_\mt{P}^{-1} (dt+ \omega)] \,.
\ee
It is now straightforward to evaluate \eqn{komar}, where the
integral is over an $S^3$ that lies at constant $\rho$ and is
parametrized by $\theta, \psi, \phi$, and the orientation is
$\epsilon_{tzr\theta\phi\psi}=+1$. The result is
\be
\calk(\rho)
= \frac{\pi^2 R_z^2}{2G}\left[Q_\mt{P} - \left(\frac{2G}{\pi^2
R_z}\right)^2\frac{J^2}{\rho^4 \, H_1 H_5}\right] \,.
\ee
Thus at $\rho \rightarrow \infty$ we recover the ADM result,
$\calk_\infty=N_\mt{P}$, whereas at the horizon we
find $\calk_\mathrm{hor}=\hat{q}_0$.

\subsubsection*{Black ring}

This is a five-dimensional black hole with horizon topology $S^1
\times S^2$. For a detailed study we refer the reader to \cite{EEMR2},
whose notation we follow. The seven independent parameters of the
solution can be taken to be the three conserved charges $Q_i$,
three dipoles $q_i$, and a radius $R$. In the M-theory
description, $N_i= (\pi/4G)^{2/3}Q_i$ and
$n_i=(\pi/4G)^{1/3}q_i$ are the numbers of constituent M2- and
M5-branes, respectively; as in \cite{EEMR2}, we use conventions
such that they are non-negative.
Regularity and absence of pathologies of the solution requires
$N_1\geq n_2n_3$ (and cyclic permutations thereof).

The Einstein-frame metric takes the form
\be
ds^2 = - f^{-2/3} (dt + \omega)^2 + f^{1/3} d{\bf x}_{\it 4}^2
\,,
\ee
where $f=H_1 H_2 H_3$,
\be
\label{eqn:flatspace}
d{\bf x}_{\it 4}^2 = \frac{R^2}{(x-y)^2} \left[ \frac{dy^2}{y^2-1}
+ (y^2-1)d\psi^2 +\frac{dx^2}{1-x^2}+(1-x^2)d\phi^2 \right]
\label{base}
\,,
\ee
\be
H_1 = 1 + \frac{Q_1 - q_2 q_3 }{2R^2} (x-y)
- \frac{q_2 q_3}{4 R^2} (x^2 - y^2) \,,
\label{eqn:Hi}
\ee
and $H_2$ and $H_3$ are given by obvious permutations. The
rotation one-form is $\omega = \omega_\phi d\phi + \omega_\psi
d\psi$, where
\bea
 \omega_\phi &=& - \frac{1}{8 R^2} (1-x^2) \left[
 q_1 Q_1 + q_2 Q_2 + q_3 Q_3 - q_1 q_2 q_3
 \left( 3 + x + y \right) \right] \,, \nn
 \omega_\psi &=& \frac{1}{2}(q_1 + q_2 + q_3) (1+y)
 + \frac{y^2-1}{1-x^2} \,\omega_\phi \,.
\label{eqn:omegas}
\eea
The coordinates $\psi$ and $\phi$ are polar angles in the plane of
the ring and in the plane orthogonal to the ring, respectively,
whereas $y$ measures the distance to the ring. Surfaces of
constant $y$ have topology $S^1 \times S^2$ and are concentric
with the ring. The horizon lies at $y \rightarrow -\infty$, and
asymptotic infinity at $y \rightarrow -1$. The integrally
quantized ADM angular momentum in the plane of the ring is
\be
J_\psi =  \left( \frac{\pi}{4G} \right)^{2/3} R^2 \left( n_1 + n_2
+ n_3 \right) + \half \left( N_1 n_1  + N_2 n_2 + N_3 n_3 -n_1n_2n_3 \right)
\,.
\ee
Note that in our conventions this is non-negative. The entropy is
given by \eqn{cardy} with $c=6n_1n_2n_3$ and
\be
\hat{q}_0 = -J_\psi + \frac{n_1n_2n_3}{4} +
\half \left( \frac{N_1 N_2}{n_3} + \frac{N_2 N_3}{n_1} +
\frac{N_1 N_3}{n_2} \right) - \frac{1}{4 n_1n_2n_3} \Big(
(N_1  n_1)^2 + (N_2 n_2)^2 + (N_3 n_3)^2 \Big) \,.
\label{qhat}
\ee
Regularity of the ring horizon requires $\qhat >0$.

The dual CFT lives on a circle that may be identified with the
$S^1$ factor of the ring horizon \cite{CGMS}, so we are interested
in the Komar integral \reef{main} associated to the Killing field
$\partial_\psi$, whose dual one-form is
\be
\ell = - f^{-2/3} \omega_\psi \, (dt+\omega) +
\frac{f^{1/3} R^2 (y^2 -1)}{(x-y)^2} \, d\psi \,.
\ee
The surface of integration is a two-sphere that links the ring
horizon; this lies at constant $t,y,\psi$ and is parametrized by
$x,\phi$. The value of the integral for arbitrary values of $y$ is rather
complicated and hardly illuminating, but it simplifies in the two
limits of interest. First,
the result at infinity (\ie in the limit $y \rightarrow -1$) is
$\calk_\infty = -J_\psi$. Incidentally, note that we obtain the
ADM angular momentum despite the fact that we are integrating over
a two-sphere. This is because (as may be verified by direct
calculation)
\begin{equation}
\int_{S^2} *(\ell \wedge d\ell)
=\frac{1}{2\pi}\int_{S^1 \times S^2} *d\ell
=\frac{1}{2\pi}\int_{S^3} *d\ell \,,
\end{equation}
where the $S^3$ is a three-sphere at infinity. The last equality
holds because the $S^1 \times S^2$ and the $S^3$ surfaces differ
only in the `axis' $y=-1$,
on which the form $* d\ell$ vanishes
identically.

On the other hand, if we evaluate the integral at the horizon (\ie
in the limit $y \rightarrow -\infty$) we get $\calk_\mathrm{hor} =
\qhat$. This is as expected, since it was shown in \cite{EEMR1} that the
near-horizon solution  is of the form \reef{nearbtz}.

As we saw above, both $J_\psi$ and $\qhat$ are non-negative. It is
interesting to note that $\qhat$ may be smaller than, equal to or
larger than $J_\psi$. Indeed, $\qhat$ may be made arbitrarily
small by keeping the charges and dipoles fixed and increasing $R$.
In this limit $J_\psi$ remains non-zero, so
$J_\psi > \qhat$. Reciprocally, we may fix $q_i$ and $R$ and
increase $Q_i$. Since $\qhat$ grows like $Q^2$ and $J_\psi$ grows
like $Q$, for large enough $Q_i$ we have $\qhat > J_\psi$.

It is interesting to note that, although the overall sign for $\calk$
depends on conventions (ours is such that it yields $+\qhat$ at the
horizon), we find that $\calk$ assigns opposite signs to $\qhat$ and
$J_\psi$. This is not to say that the horizon is counter-rotating
relative to infinity, since it actually is static \cite{EEMR1}. A
similar difference between the sign of the Komar integral at
infinity and at the horizon was found for the BMPV black hole in
\cite{GMT}. In the present case, it is presumably related to the fact that
in the CFT calculation of $\qhat$ in \cite{CGMS}, one must
identify the total momentum $q_0$ with $- J_\psi$.

\subsubsection*{Solutions with varying charge density}

Supersymmetric, ring-like solutions with an inhomogeneous,
finite-area, singular horizon of topology $S^1 \times S^2$ were
constructed in \cite{BWW}. They are characterized by the fact that
the M2-brane charge densities are not constant along the
$\psi$-direction but are specified by arbitrary functions,
so $\partial_\psi$ is no longer an isometry. The `horizon'
fails to be a regular one \cite{HR}, but an entropy can still
ascribed to these solutions. This takes the form
\be
S = \int_0^{2\pi} d\psi \, \sqrt{\frac{c}{6} \, \qhat(\psi)}\,,
\ee
where $c=6 n_1 n_2 n_3$,
\be
\qhat(\psi) = \frac{1}{n_1 n_2 n_3}
\left( \frac{\pi R}{G} \right)^2 \hat{M}(\psi) \,,
\ee
and $\hat{M}(\psi)$ is a function of the local M2-brane charge
densities on the ring \cite{BWW}. Thus
$\qhat(\psi)$ represents a local oscillator level on the
ring. We have verified by direct calculation that this is exactly
measured by the Komar integral
\eqn{komar} evaluated on a two-sphere at the horizon that lies at the
corresponding value of $\psi$, that is
$\calk_\mt{hor} (\psi) = \qhat(\psi)$. Presumably an analogous result
will hold for the inhomogeneous black string `horizons' with
travelling waves of \cite{HM}.

\subsubsection*{Black ring in Taub-NUT}

Black rings with Taub-NUT asymptotic boundary conditions have been
constructed in \cite{EEMR3,Strominger,BK-TN}. For these solutions the
ring wraps a circle direction that, at asymptotic infinity, has
finite radius. Hence the solution can be seen as carrying linear
momentum in this direction.

The dual CFT can again be thought of as living on the $S^1$ factor
of the horizon. Computing the integral \reef{komar}, the result for a
two-sphere at infinity is
$\calk_\infty = -N_\mt{P}$, whereas at the horizon we find $\calk
= \qhat$, as expected in both cases.

\subsubsection*{4D black holes}

The familiar supersymmetric four-charge 4D black hole, obtained by
reduction of a five-dimensional black string, provides a rather trivial
example for our calculations, since in this case $\qhat$ coincides with
the ADM momentum. Less trivial examples are obtained using the 5D black
ring to obtain 4D black holes. When the asymptotically flat black ring
has very large $R$, it becomes a supersymmetric five-dimensional black
string \cite{bena}, which can be reduced to a four-dimensional black
hole. Since this is a particular limit of the black ring, it is clear
that $\calk$ will also perform correctly the dual of the CFT calculation
of the oscillator level. The black ring in Taub-NUT provides in fact a
more general example, since it directly reduces to a four-dimensional
black hole configuration. Finally, it should be possible to lift to five
dimensions the generating black hole solution of \cite{BT} to exhibit
the effective string direction. Given that it admits a 2D CFT
dual, the near-horizon structure must be of the form \reef{nearbtz}, and
so it must also conform to our general analysis.

\section{Discussion}
\label{discussion}

The main result in this paper is the simple observation that a
Komar integral is the function that naturally interpolates between
the values of the momentum at the horizon and at infinity. This
has a nice interpretation for supersymmetric black holes with a
dual 2D CFT, since their near-horizon structure is such that this
interpolation amounts to the computation of the oscillator level.
Such a calculation can be rather non-trivial in the CFT, as the
example of supersymmetric black rings illustrates, but the
integral \reef{komar} performs it neatly. Black rings also show
that in order to identify the relevant parameter near the horizon,
the Komar integral must be performed on surfaces adapted to the
horizon topology, namely $S^1\times S^2$, as opposed to the
$S^3$ that must be used for spherical black holes.

The above investigations suggest that other Komar integrals may be
relevant to the microscopic study of black holes. An interesting
example is that of the Komar integral associated to rotations in
the plane orthogonal to the ring circle for black rings, \ie
rotations along the angle
$\phi$ of the $S^2$. It is puzzling that the black ring possesses
a non-zero angular momentum in this plane, yet there seems to be
no trace of it in the microscopic description
\cite{EEMR2, CGMS}. It was suggested in \cite{EEMR2} that this
angular momentum is not carried by the ring itself but is in fact
entirely generated by crossed electric and magnetic supergravity
fields outside the black ring horizon. A necessary condition for
this interpretation to be consistent is that the Komar integral
associated to $m=\partial_\phi$,
\be\label{phiint}
\frac{1}{8G}\int_{S^2} *(\ell \wedge dm)\,,
\ee
vanish when evaluated at the ring horizon. We have verified by
direct calculation that this is indeed true. At infinity, the same
integral yields the corresponding ADM quantity, $-J_\phi$, as
expected.

Although this result is certainly suggestive, one should take it
with some caution, since the use of Komar integrals for localizing
a physical magnitude is not without pitfalls. For example, the
Komar integral associated to the Killing vector $\partial_t$ is
zero on the horizon of an extremal black hole (due essentially to
the vanishing surface gravity), but it would be doubtful to
interpret this as saying that the mass of the microscopic
constituents of extremal black holes is zero. Instead, it seems
more appropriate to conjecture that the Komar integrals at the
horizon measure only those quantum numbers that are associated to
a degeneracy of states of the dual CFT. All contributions to the
Komar integral that do not add to the degeneracy are subtracted
off when going from infinity to the horizon.

Incidentally, note that our results for the Komar integral at
infinity for black rings suggest that it is natural to identify
the total momentum of the CFT with $-J_\psi$ \cite{CGMS}. This
differs from the identification in \cite{BK} of the CFT momentum
as $J_\psi
-J_\phi$, which would instead require considering the difference
between
\reef{komar} and \reef{phiint}.

We have only considered supersymmetric black holes, but it is
natural to ask whether Komar integrals may also be useful to
understand the microscopic description of non-supersymmetric black
holes \cite{BLMPSV} and black rings \cite{previous,RE,EEF}, or,
even further, of configurations without horizons such as
supertubes \cite{MT} or the recently constructed non-singular
solutions of \cite{nonsing}. For non-singular, horizonless
solutions the integration surface for the Komar integral can be
smoothly shrunk to vanishing size and hence should yield zero, in
line with the fact that these solutions should have no entropy, as
they correspond to individual microstates.

The question seems less trivial for near-supersymmetric black
holes. These can also be described within a dual 2D CFT, and it is
natural to ask whether there exist Komar integrals that separately
compute their left- and right-moving oscillator levels. Since in
these cases the mass is an independent parameter, it is natural to
consider Komar integrals associated to $\partial_t$ in order to
identify the (non-zero-mode) energy above the CFT ground state. It
is not difficult to find a combination \textit{ad hoc} that yields
correct results for the oscillator levels of the near-extremal
five-dimensional rotating black hole solution \cite{BLMPSV}, whose
decoupling limit is well known to be BTZ $\times S^3$ (where in
general the $S^3$ angular coordinates are shifted by the BTZ time
and space directions) \cite{CL}. Let $k$ be the Killing vector
that generates time translations, normalized in such a way that,
at infinity, $k+\ell$ is a null vector. We then consider the Komar
integral
\be
\cale =-\frac{1}{16G}\int_{S}*(k \wedge d\ell+\ell \wedge d k)\,.
\ee
At the horizon, this measures the excitation energy $\cale$ above
the ground state of these solutions, since
\be
\hat{q}_{0L}= \frac{\cale_\mt{hor}+\calk_\mt{hor}}{2}\,,
\quad \hat{q}_{0R}=\frac{\cale\mt{hor}-\calk_\mt{hor}}{2}\,,
\ee
reproduce the correct left and right oscillator levels for the
near-extremal solutions of \cite{BLMPSV}. It is easy to see that
in the extremal case $\cale=\calk$.

However, this prescription is not well defined for black rings
(not even supersymmetric ones).\footnote{Note also that the most
general non-supersymmetric black rings known to date \cite{EEF}
cannot describe the near-supersymmetric excitations above
\textit{regular} BPS black rings. Also, the connection between
magnitudes near the horizon and at infinity is a main missing link
in the microscopic picture of near-supersymmetric rings in
\cite{larsen}.} Even for the near-extremal five-dimensional black
hole, we lack a generic interpretation for $\cale$ near infinity,
as well as an argument analogous to the one we have given in
section \ref{general} near the horizon (in the sense of
\cite{Reall}, and not as decoupling limit). This suggests that
some of the difficulties may be associated to the different
near-horizon structure of non-supersymmetric black holes. It is
tempting to conjecture that the deeper reason is that this
structure is not controlled by the attractor mechanism
\cite{attr}, although at present it is not clear to us whether
there is any significant connection between the evolution of
$\calk$ and the attractor flow. Both involve the evolution of a
certain function from asymptotic infinity towards a specific value
at the horizon, and indeed, the same generic argument that the
entropy, being a quantized number, must be independent of
asymptotic moduli, appears to imply as well moduli-independence of
our Komar integral at the horizon. If such a connection does
actually exist, it might be useful in particular for
supersymmetric black rings, for which the attractor mechanism
presents peculiarities which have been discussed in
\cite{KL}.

Finally, it would be interesting to study whether our results
can somehow be useful as well for black holes with a dual $D>2$ CFT.

 \medskip
\section*{Acknowledgements}
\noindent
We thank Don Marolf for discussions. RE is grateful to the
Perimeter Institute for warm hospitality. RE was supported in part
by CICYT FPA 2004-04582-C02-02 and European Comission FP6 program
MRTN-CT-2004-005104.


 \end{document}